# Energies for cyclic and acyclic aggregations of adamantane and diamantane units sharing vertices, edges, or six-membered rings

Alexandru T. Balaban, Debojit Bhattacharya, Douglas J. Klein and Yenni P. Ortiz

Texas A&M University at Galveston, 200 Seawolf Parkway, Galveston, TX 77553, USA

## Abstract

Diamondoids are hydrocarbons having a carbon scaffold comprised from polymer-like composites of adamantane cages. The present paper describes computed total energies and "SWB-tension" energies (often referred to as "strain" energies) for species having *n* adamantane or diamantane units sharing pairwise: one carbon atom (spiro-[*n*]adamantane or spiro-[*n*]diamantane); one C—C bond (one-bond-sharing-[*n*]adamantane or one-bond-sharing-[*n*]diamantane); or one chair-shaped hexagon of carbon atoms (1234-helical-cata-[*n*]diamantanes). Each of the five investigated polymer-like types is considered either as an acyclic or a cyclic chain of adamantane- or diamantane-unit cages. With increasing *n* values, SWB-tension energies for acyclic aggregates are found to increase linearly, while the net SWB-tension energies of cyclic aggregates often go thru a minimum at a suitable value of $n$. In all five cases, a limiting common energy per unit ($E/n$) is found to be approached by both cyclic and acyclic chains as $n \to \infty$, as revealed from plots of $E/n$ versus $1/n$ for acyclic chains and of $E/n$ versus $1/n^2$ for cyclic chains.

## 1. Introduction

There exist two main carbon allotropes found in nature: $sp^2$-hybridized graphite, black, electrically-conducting, and thermodynamically more stable under normal conditions of pressure and temperature, and the $sp^3$-hybridized denser, transparent, electrically insulating and hard diamond. Graphite can be converted industrially into synthetic diamond at about 54 kbar (5.4 GPa) and 1700 K in molten nickel.[1,2] Chemical vapor deposition of diamond-like films at high temperatures around 900º C is now being used commercially for obtaining die cutters for nonferrous materials.[3]

In graphene, the hydrocarbon corresponding to a unit cell of high point-group symmetry is benzene $(CH)_6$, and in diamond a correspondent high-symmetry "reduced" cell is adamantane $C_{10}H_{16}$. Adamantane has the partition formula $(CH)_4(CH_2)_6$ when one considers the degrees of the molecular graph representing the carbon scaffold. Analogously, diamantane which consists of two adamantane units (cages) sharing a common 6-membered ring, has molecular formula $C_{14}H_{20}$ and partition formula $(CH)_8(CH_2)_6$.

Diamondoid hydrocarbons (henceforth diamondoids for short) have acquired sudden prominence since Dahl and Carlson published their discovery of polymantanes in petroleum. [4] As an isolated event, adamantane had initially been first identified in petroleum by Landa,[5] then prepared by Prelog in an elaborate synthesis, [6] and finally obtained serendipitously by Schleyer via Lewis-acid-catalyzed multistep rearrangement from a polycyclic isomer.[7] In a series of papers, using Dahl and Carlson's diamondoids as starting materials, Schreiner and coworkers described many potential applications of diamondoids for biomedicine and electronics,[8-12] while Drexler earlier proposed diamond nanostructures to build nanomachines.[13] Similarly to the conversion of graphite into diamond. Dahl, Carlson, Schreiner and their coworkers showed [14] that diamantane-4,9-dicarboxyxlic acid becomes decarboxylated and polymerized by the "capillary" force exerted inside carbon nanotubes of sufficiently large diameter (around 1.0 – 1.3 nm), thereby resulting in long zigzag catamantanes. It is also possible to convert benzene, at room temperature under high pressure (20 GPa), into carbon nanothreads with diamondoid structure, using a diamond anvil,[15] as hinted at earlier by Drickamer.[16]

Adamantane, diamantane and triamantane contain one, two, and three adamantane units (cages or cells), respectively, and are unique isomers (Fig. 1), but tetramantane and higher diamondoids have more than one isomer. Using the analogy with benzenoid hydrocarbons, [17,18] Balaban and Schleyer [19] devised a simple system of encoding structures of diamondoids based on the concept of dualists (or inner dual graphs). Centers of adamantane units are the vertices of dualists; edges of dualists connect vertices corresponding to adamantane units sharing six-membered rings of carbon atoms. Also by analogy with benzenoids, diamondoids are classified into *catamantanes* when their dualists are acyclic, *perimantanes* when their dualists have six-membered rings, and *coronamantanes* when the adamantane units viewed as solids end up giving a solid of genus 1 – and yet higher-genus structures are conceivable. It should be stressed that unlike graphs, the geometry of dualists matters (edges are straight lines with definite lengths and angles). This scheme [19] reduces the specification of polymantanes to that of specifying of saturated hydrocarbons conformations embeddable on the diamond lattice – though this encompasses the specification of polymantanes, the utility is that the conformations resulting are for much smaller saturated hydrocarbons. A general (related) scheme of specification is described elsewhere.[20]

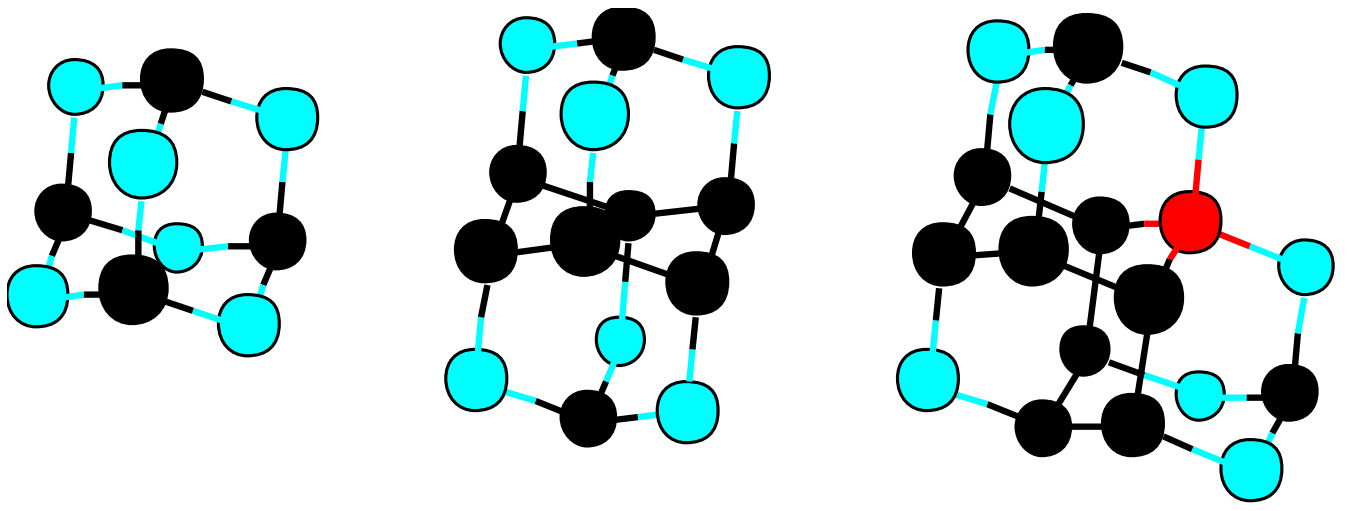

Fig. 1. Carbon scaffolds of adamantane ($C_{10}H_{16}$), diamantane ($C_{14}H_{20}$), and triamantane ($C_{18}H_{24}$); the color code has $CH_2$ groups in blue, CH groups in black, and tetra-connected C atoms in red.

By assigning numbers 1, 2, 3, and 4 to the bond directions around an $sp^3$-hybridized carbon atom, a simple way of encoding diamondoid structures was devised on the basis of selecting the smallest number among all listings of bond directions along the longest path of the dualist.[19] For tetramantanes, there are three possible constitutional isomers namely the non-branched zig-zag [121]tetramantane, the branched [1(2)3]tetramantane, and the chiral [123]tetramantane. Since the latter structure corresponds to two enantiomers, there are four stable isomeric substances called tetramantane. However, for specifying substituent positions, the IUPAC (von Baeyer) nomenclature of catamantanes has to be used.[21,22] A similar approach was also used for the IUPAC nomenclature of certain perimantanes.[22]

In a recent paper,[23] one of the present authors described various ways in which several hydrogen-depleted molecular graphs of adamantane may share vertices, edges or faces (shared hexagons of carbon atoms). Klein and coworkers [24] described a construction to generate novel super-adamantane structures based on the simplest connection via C–C bonds between tertiary carbon atoms of different adamantane units. In the paper mentioned above [23] various ways to connect adamantane units into acyclic or cyclic arrays of spiro-adamantanes and torus-like poly-adamantanes were described. There are also various ways in which adamantane units can share one or two C–C bonds according to the topological distances between them. For two adamantane rings sharing 1, 2, 3, or 6 carbon atoms, Schleyer and coworkers proposed the names [1]diadamantane, [2]diadamantane, [3]diadamantane, and [6]diadamantane, respectively,[25] but such names are not easily generalized to longer acyclic or cyclic assemblies of adamantane units; neither can they be adapted to systems composed of diamantane units. Except for the third term, the other three hydrocarbons have all been prepared. [26-35]

In the present paper we display results of calculations for chains of several classes of diamondoid structures, starting with *spiro-[n]adamantanes* when a vertex is shared pairwise by *n* adamantane units, or *spiro-[n]diamantanes* when a vertex is shared pairwise by *n* diamantane units. Spiro[adamantane-2,2'-adamantane] (which we call for simplicity spiro-[2]adamantane) is the simplest spiro-adamantane.[26,27] Thermal isomerizations of spiro-adamantanes were reported.[28,29] We confine our investigation to spiranic structures that involve sharing carbon atoms from $CH_2$ groups which are farthest apart from each other. We describe also in the present paper chains involving one-bond-sharing poly-adamanatanes and poly-diamantanes; finally we describe in this paper chains of helical cata-diamantanes, which share chair-shaped hexagons of carbon atoms. Throughout investigate two forms of arranging monomer units: either as open chains termed *acyclic*, or as chains formed into a ring, and termed *cyclic.*

Generally in the cubic diamond network and polymantanes, fused adamantane units are each of tetrahedral symmetry $T_d$ with tetrahedral directions pointing alternatively in opposite directions. Diamantane units have a single preferred direction, along the longer axis of this molecule, whence distal (*i.e.* farthest apart) groups are more uniquely defined than for adamantane, as shown in Table 1. For distinguishing among distances between bonds in adamantane or diamantane units, the bonds are labeled with letters as seen in Fig. 2.

Table 1. Topological distances between shared features in adamantane and diamantane units

| Most distant shared features | Adamantane | Diamantane ( involving apical CH group) |
|---|---|---|
| $CH_2$ groups from $CH_2$ group | 1 | 1 |
| CH groups from CH group | 3 | 1 |
| C–C bonds from C–C bond | 4* | 1 |
| Hexagon from hexagon | 3 | 1 |

*Two of these (**d**) are parallel to edge **a** in common hexagons, and two (**d'**) are not parallel, having no common hexagons.

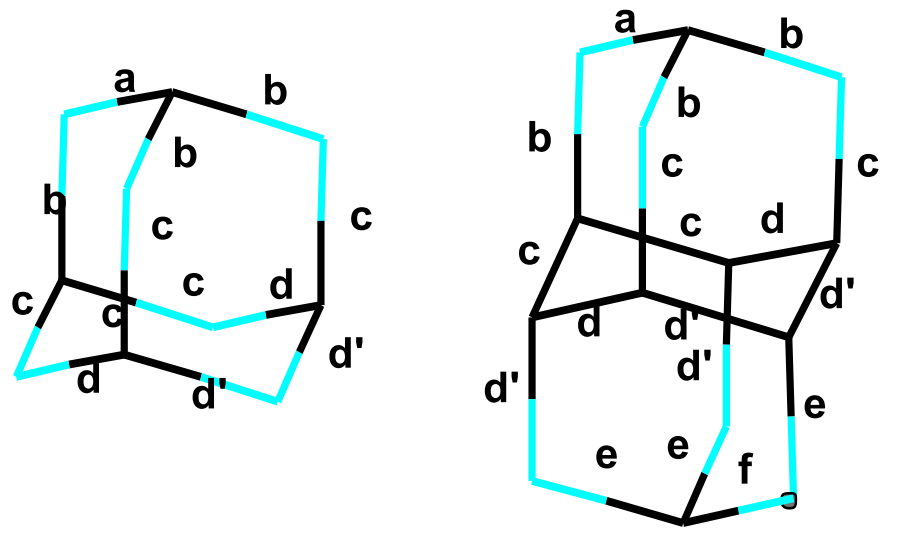

Fig. 2. Edge labeling by letters in adamantane and diamantane

Consequently, along with the structures based on adamantane units examined in the previously mentioned paper,[19] a few other structures based on diamantane units are also considered. Two apical vertices of the carbon scaffold in diamantane (tertiary CH groups) stand out as being at maximum distance, but they would only be involved in necklace-like structures, which are not discussed here; however, three pairs of secondary $CH_2$ groups also are at larger distances than the other $CH_2$ groups, and such groups become shared vertices in spiro-diamantanes. The three pairs of parallel C–CH bonds in diamantane units involving the pair of apical tertiary CH groups, having the highest inter-bond distance, will constitute the one-edge-shared chains and rings. Finally, we examine structures in which diamantane units share their most distant 6-membered rings. Results of "tension" (or "strain") energy calculations will be presented in the following in order to investigate dependences on the numbers of adamantane or diamantane units. Thus there are five classes of diamondoid cyclic and acyclic chains to be examined in the present paper. Vertices (carbon atoms) and edges (bonds) in each of these classes are shown by graphs of hydrogen-depleted structures in Fig. 3.

Fig. 3. Upper row, from left to right: adamantane units sharing common $CH_2$ groups situated in distal positions and yielding spiranic chains; and adamantane units sharing one bond. Lower row, from left to right: diamantane units sharing apical $CH_2$ groups with neighboring units and yielding spiranic chains; sharing one bond; and sharing a 6-membered ring to the lower left and upper right with neighboring diamantane units. Shared $CH_2$ and CH groups are indicated in red.

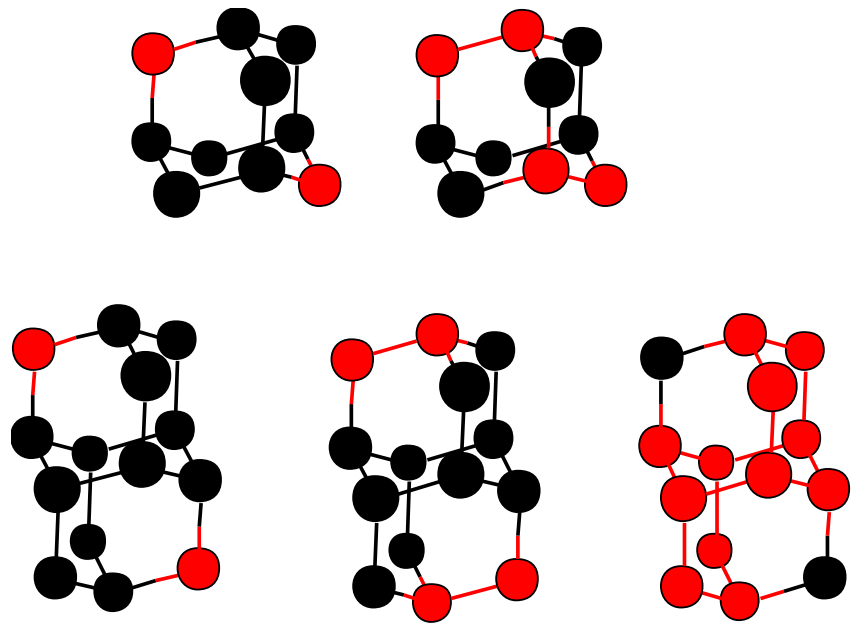

In 1970, Schleyer, Williams and Blanchard [36] presented a "group-contribution-like" scheme for estimation of heats of formation with the higher-order (longer-range) contributions being identified to what was termed "strain". Within this scheme they describe diamondoids as not being "strain-free", despite the fact that bond angles are tetrahedral, bond lengths are normal, and conformations are staggered. But this nomenclature is faulty in the sense that "strain" properly relates to geometry, while "stress" properly relates to force. Thus in such a proper nomenclature, it may be viewed that they maintain that these diamondoid structures are stressed while being strain-free – this stress being viewed to originate in various non-neighbor carbon-carbon repulsions. But the situation is worse in that for our cyclic-chain aggregations, there is a distortion in geometry (such as is more conventionally termed a strain), and thereby gives rise to a (proper) strain energy (quadratic in this strain), so that we would be faced with a strain-energy contribution to Schleyer-Williams-Blanchard's "strain energy". Rather than perpetuate such a problematic nomenclature we henceforth refer to Schleyer-Williams-Blanchard "strain" energy as the *SWB-tension* energy. In adamantane this SWB-tension energy amounts to about 6 kcal/mol, and in homologous diamondoids it increases with the number of adamantane units. It may also be noted that there are other counter choices as to how tension (or "strain") energies should be referenced,[37-39] avoiding the assignment of stress or tension in adamantane or other diamondoids. But here we utilize the SWB-tension energy delivered by the MM2 package we use. These energies still eliminate the large contributions to the total energy.

A note about terminology needs to be added. Constituent building blocks of diamondoid hydrocarbons to be discussed below are damantane or diamantane units or cells, and the result is a covalently-bonded aggregate or polymer, shown as a hydrogen-depleted structure.

## 2. Expected Behaviors

Results for the various polymers treated can be viewed in a simple manner to have different properties, such as total energies or SWB-tension energies, approximated in an "additive" manner. For instance, with the presumption of a SWB-tension energy $\varepsilon_{\mathrm{int}}$ for an internal unit and another energy $\varepsilon_{\mathrm{end}}$ for an end (terminal) unit which is less tensed, one might imagine that the total net SWB-tension energy is:

$$E_n \cong \varepsilon_{end} + (n-2)\varepsilon_{in} + \varepsilon_{end} \tag{1}$$

As a consequence one obtains eq. (2)

$$E_n / n \cong \varepsilon_{in} + \frac{2(\varepsilon_{end} - \varepsilon_{in})}{n} \tag{2}$$

so that a plot of $E_n/n$ versus $1/n$ would be anticipated to be asymptotically linear, with negative slope (when $\varepsilon_{\mathrm{in}} > \varepsilon_{\mathrm{end}}$) and an intercept equal to $\varepsilon_{in}$. Indeed there is evidence [40] that such a functional form is highly accurate, with plausibly exponentially small corrections. And in Fig. 4 we see that this expectation is borne out, with a correlation coefficient > 0.9999 – and indeed such a result continues for our further examples of open chain structures.

For cyclic spiro-[$n$]adamantane structures, within our "additive" analysis in terms of adamantane units, we now have $n$ equivalent units, each of which (rather than having two less tensed end-units with energy $\varepsilon_{end}$) has the same contribution $\varepsilon_n'$ (if there is $n$-fold cyclic symmetry). But as this $\varepsilon_n'$ is geometrically distorted, it manifests curvature strain, and should generally be $> \varepsilon_n$. The curvature strain (in bond lengths and bond angles) per unit is inversely proportional to the radius of curvature which in turn is proportional to $n$. And the curvature-mediated stress energy should naturally be proportional to the square of this geometric curvature strain. Thus the net SWB-tension energy $E_n'$ should be $n \cdot (\varepsilon_{in}' + \gamma / n^2)$ with $\gamma$ a (positive) curvature parameter. Consequently

$$E_n' / n = \varepsilon_n' \cong \varepsilon_{in} + \frac{\gamma}{n^2} \tag{3}$$

so that a plot of $E_n' / n$ versus $1/n^2$ is anticipated to be linear. The (geometric) strain $\sim 1/n$ measures deviations from the ideal unstressed unit with energy $\varepsilon_{in}$, and so it is imagined that there should be higher order corrections in $1/n$ (say $\sim 1/n^3$ or $\sim 1/n^4$). Thus we can anticipate our cyclic species might manifest a slower convergence – as indeed turns out to be the case as gauged by our generally found lower correleation coefficients.

Since the two contributions to the net SWB-tension energy $E_n'$ (for an $n$-unit cycle) are antagonistic, one can address a question as to an extremum. To do so one can set the derivate of $E_n'$ with respect to $n$ to be $\approx 0$, to give

$$n_{\min} \approx \sqrt{\gamma / \varepsilon_{in}} \tag{4}$$

the value of $n$ for the occurrence of the minimum net SWB-tension energy:

$$\min_n E_n' \approx \varepsilon_{in} n_{\min} + \gamma / n_{\min} \approx 2\sqrt{\varepsilon_{in}\gamma} \tag{5}$$

Formulas (4) and (5) are based on an asymptotic expression (neglecting higher order corrections), and therefore they should only be expected to be accurate when equation (4) gives a large value for $n_{\min}$.

## 3. Computations

Molecular energies (in eV) were computed with the semiempirical program PM6, and so-called SWB-tension energies (in kcal/mol) with Allinger's molecular mechanics MM2 approach. Cyclic and acyclic diamondoid aggregates present a linear dependence versus $n$ for PM6- (and MM2-) calculated energies with correlation coefficients $R^2$ close to 1, as is discussed in the following. Of course, since the numbers of carbon and hydrogen atoms in molecular formulas depend linearly on $n$, the same correlations for these two types of energies will be found in terms of the numbers of carbon or hydrogen atoms.

For SWB-tension energies of cyclic aggregates of all five diamondoid classes display some special features. First there seems to be some difficulty in convergence toward the desired minima, perhaps involving local minima, thereby yielding a poorer correlation coefficient (less close to 1). Also the innate tension due to non-neighbor (even nearly strain-free) interactions *increases* with $n$, whereas the curvature-induced strain and stress (due to extra bond-angle and bond-length strain) should *decrease* with $n$ much as for cycloalkanes. Then from these two antagonistic effects, one expects that the SW-tension energy should go through a minimum as $n$ increases, as already discussed, and further as sometimes observed in our later (supplemental) tables showing MM2-calculated SW-tension energies (with red-colored rows identifying the $n$ value corresponding to minimal SW-tension energy).

### 3.1. Spiro-[*n*]adamantanes

When $n$ adamantane units share vertices of their carbon scaffolds pairwise, a spiro-[*n*]adamantane results. When proximal $CH_2$ groups are involved, the resulting spiro-adamantane structures may be quite diverse but are less interesting. Here we examine spiro-[*n*]adamantanes formed when uniquely defined distal (remotest) $CH_2$ groups are shared. As seen in Fig. 3, distal vertices of adamantane are uniquely defined.

The molecular formulas of acyclic and ring-shaped spiro[*n*]adamantanes are $C_{9n+1}H_{12n+4}$ and $C_{9n}H_{12n}$, respectively. The partitioned formulas are $C_{n-1}(CH)_{4n}(CH_2)_{4n+2}$ and $C_n(CH)_{4n}(CH_2)_{4n}$, respectively.

For each new $C_9H_{12}$ or $C(CH)_4(CH_2)_4$ adamantane unit the corresponding PM6-computed energy was found to increase by 1266.5 eV both for acyclic and cyclic chains. SW-tension energies computed with the MM2 program for acyclic chains of spiro-[*n*]adamantane units (Table S1) show a linear increase in terms of $n$ with an increment of 27.6 kcal/mol for each new adamantane unit. The main components of this SWB-tension energy are torsion energy and 1,4-Van der Waals energy, with stretch and bend energies as minor components. For cyclic aggregates (Table S2) the SWB-tension energy passes through a minimum when $n$ increases, and this is a general feature in all five cases examined here; the "minimum row" is colored in red in all tables for cyclic aggregates (Tables S2, S4, S6, and S10).

In Fig. 4 the plot for the spiro-cyclo[*n*]adamantane SWB-tension energy per adamantane unit ($E'_n / n$) versus $1/n^2$ reveals a reasonable agreement, with a correlation coefficient close to 1. The two asymptotic intercept values should, according to our arguments be the same, and the results in our plots are close. Because of the evident more severe stress in the cyclic case, the result for the acyclic case should presumably be more accurate.

Cyclic spiro-[$n$]adamantanes with the same orientation (without geometric torsion) between consecutive adamantane units are expected to have extra bond-angle strain. Then the associated stresses per unit are anticipated to decrease with increasing $n$. Then from these antagonistic effects, one finds (Table S1) that steric SWB-tension goes through a minimum around $n$ = 10. Our least-squares-fitted extrapolations for the SWB-tension energy and the total energy per adamantane unit (as $n \to \infty$) are quite similar for both acyclic and cyclic cases, namely 27.7 kcal/mol, and 1266.4 eV, respectively.

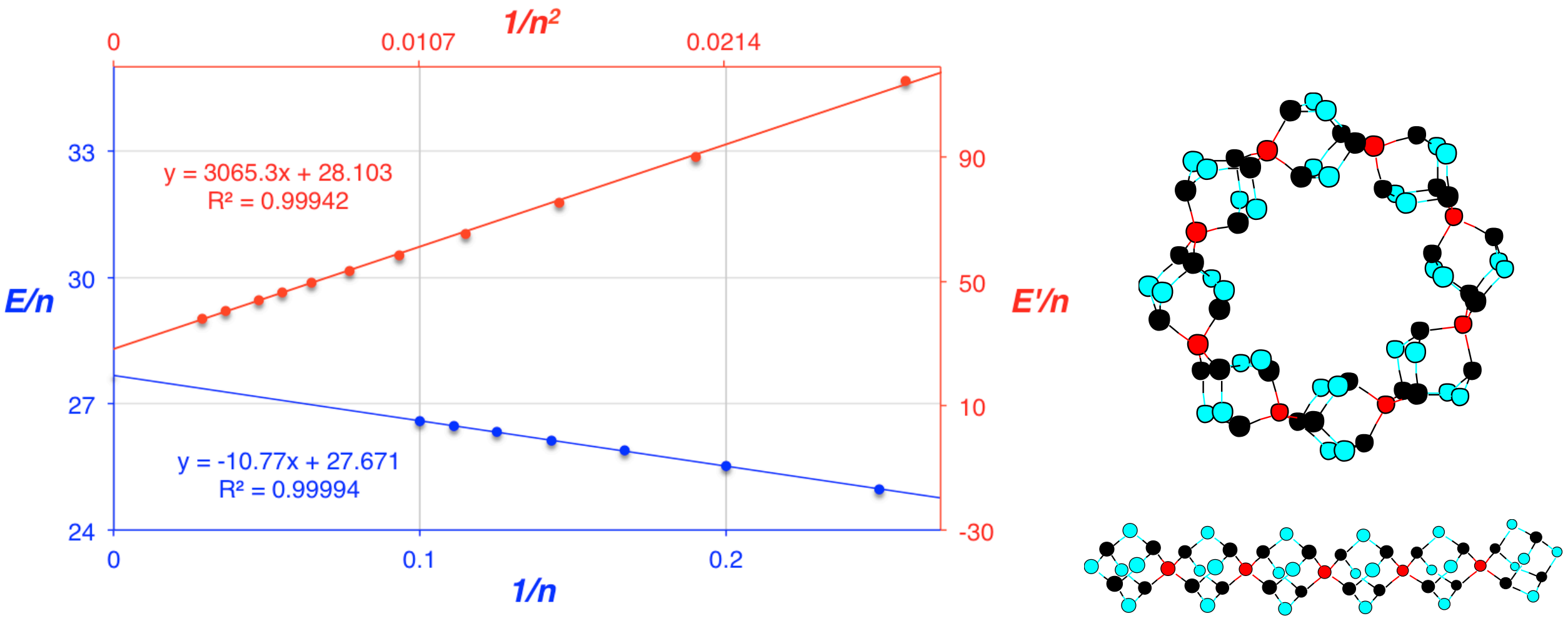

Fig 4. SWB-tension energies (in kcal/mole) for spiro-[$n$]adamantanes. Upper plot in red: cyclo-spiro[8]adamantane with energy per adamantane unit versus $n^{-2}$. Lower plot in blue: spiro[6]adamantane with energy per unit versus $n^{-1}$. For structures, the color code as in Fig. 1.

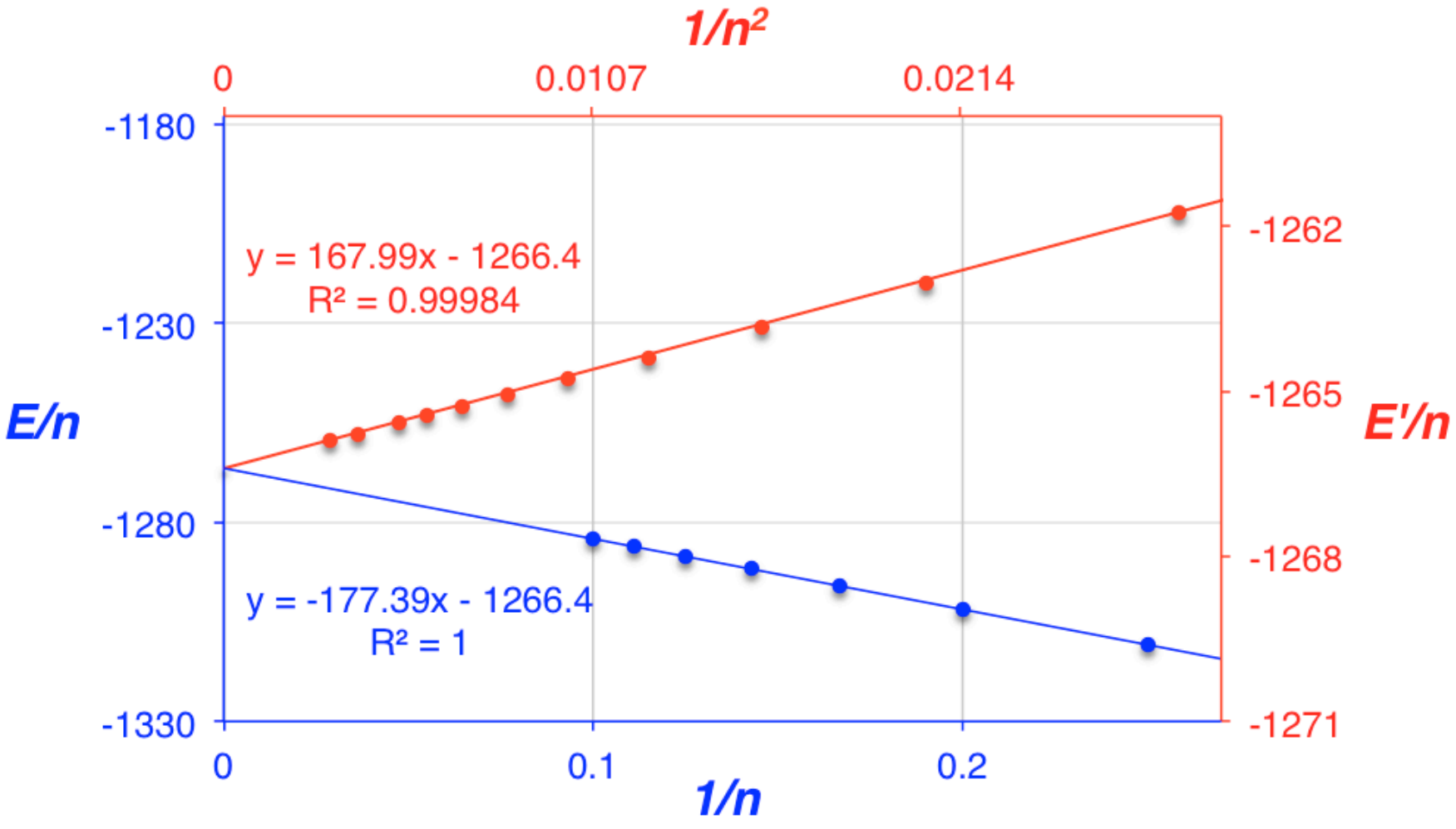

Fig. 5. Total energies (in eV) for spiro-[$n$]adamantanes. Upper plot in red: cyclo-spiro[$n$]adamantane with total energy per adamantane unit versus $n^{-2}$. Lower panel in blue: spiro[$n$]adamantane with total energy per unit versus $n^{-1}$.

### 3.2. Spiro-[$n$]diamantanes

Similar findings occur for chains of spiro-[$n$]diamantane, since again pairs of remotest apical $CH_2$ groups are uniquely defined. Acyclic and ring chains have molecular formulas $C_{13n+1}H_{16n+4}$ and $C_{13n}H_{16n}$, respectively; the partitioned formulas are $C_{n-1}(CH)_{8n}(CH_2)_{4n+2}$ and $C(CH)_{4n}(CH_2)_{4n}$, respectively. Figure 6 presents energies for spiro[5]diamantane and cyclo-spiro[8]diamantane, again following equations (2) and (3). The linear increment for the MM2 SWB-tension of acyclic diamantane chains is 36.6 kcal/mol. The total energy increase for each new $C_{13}H_{16}$ or $C(CH)_8(CH_2)_4$ diamantane unit is 1811.1 eV.

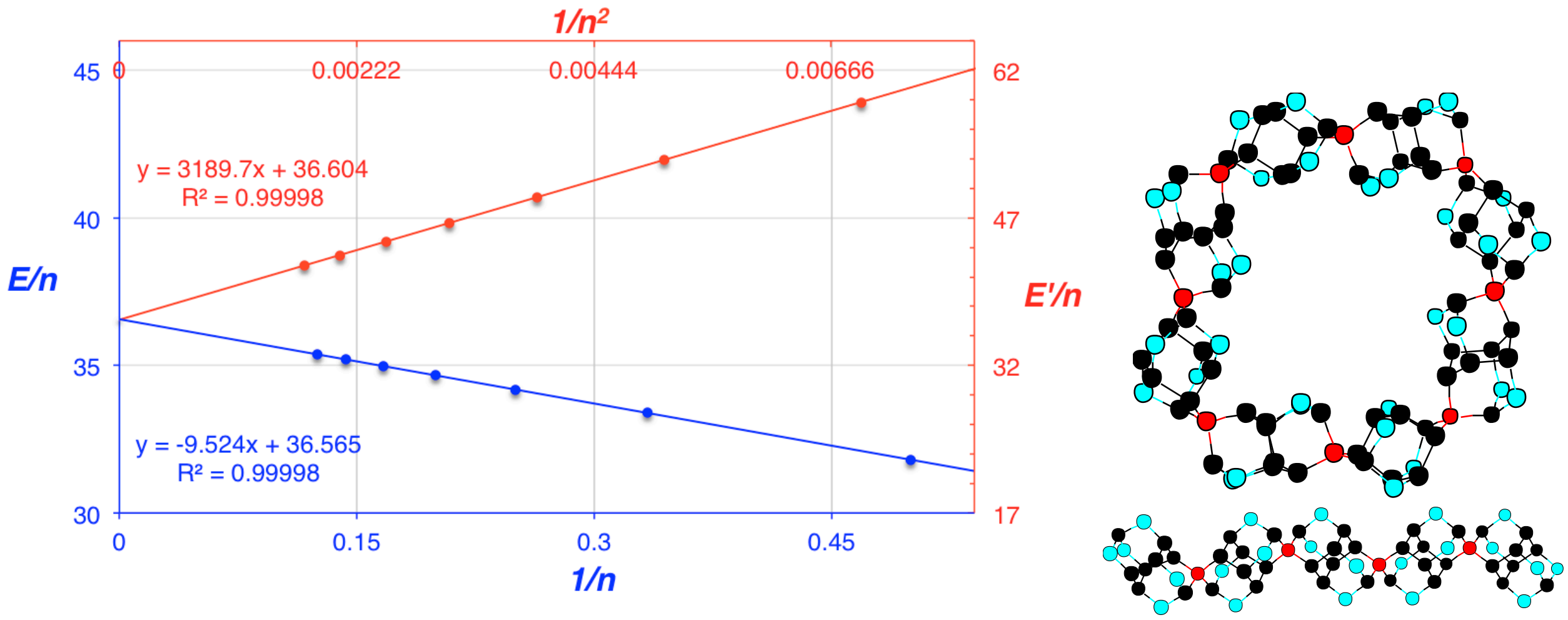

Fig 6. SWB-Tension energies (in kcal/mol) for spiro-[$n$]diamantanes. Upper plot in red: cyclo-spiro[8]diamantane and the plot of the SWB-tension energy per adamantane unit versus $n^{-2}$. Lower plot in blue: spiro[5]adamantane and the plot of the SWB-tension energy per unit versus $n^{-1}$. For structures, the color code as in Fig. 1.

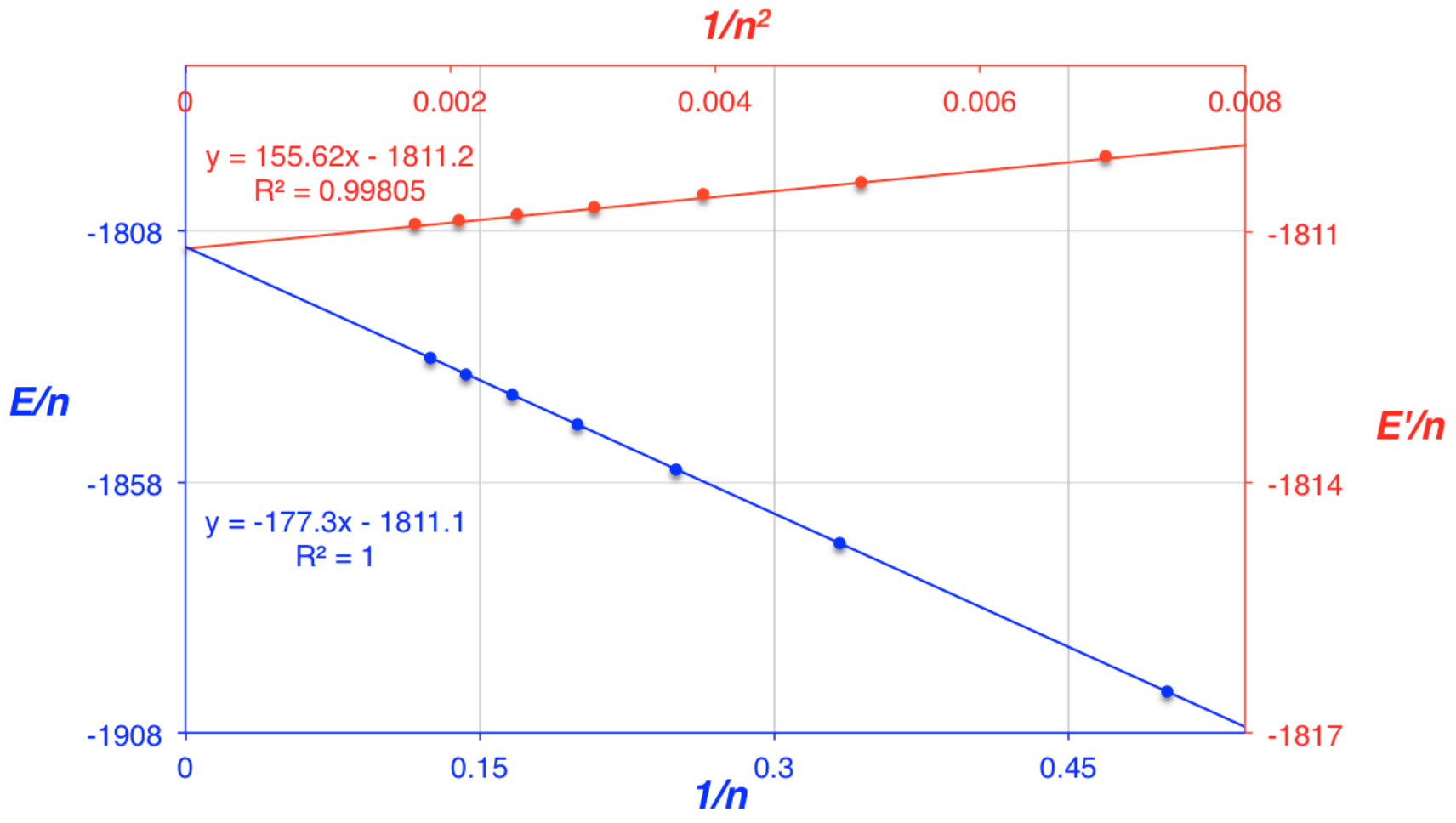

Fig. 7. Total energies (in eV) for spiro-[$n$]diamantanes. Upper plot in red: cyclo-spiro-[n]diamantanes and the plot of the total energy per diamantane unit $En^{-1}$ versus $n^{-2}$. Lower plot in blue: spiro-[$n$]diamantanes and the plot of the total energy per unit versus $n^{-1}$.

### 3.3 One-bond-sharing-[*n*]adamantanes

We examined aggregates of adamantane units sharing one bond with adjacent units (see Fig. 3) such that bonds **a** and **d’** of Fig. 2 are shared in internal units. When adamantane units share one bond to afford quasi-straight acyclic aggregates (as in Fig. 8), again the SWB-tension energy increases linearly (see Table S4). On joining the ends of such aggregates without twisting, one obtains one-bond-sharing (abbreviated as one-BS-) cyclo-[$n$]adamantanes; examples with $n$ = 11 and 24 are presented in Fig. 8.

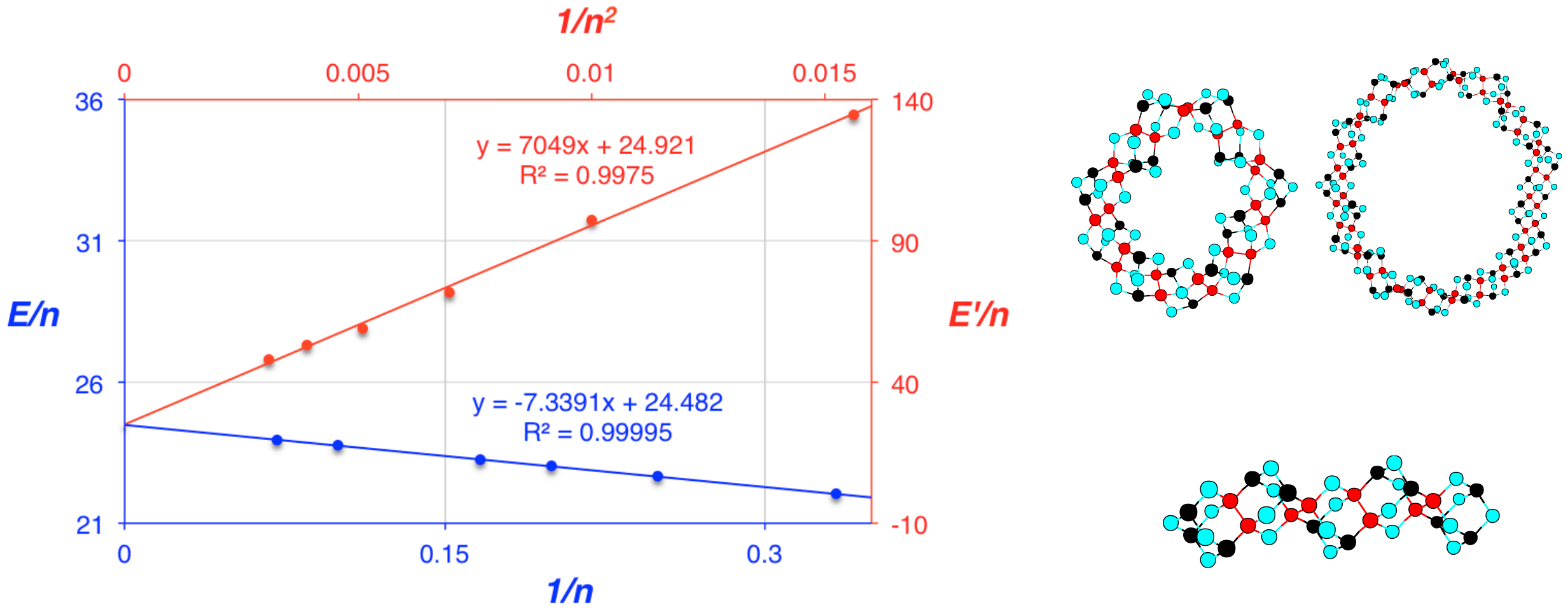

Fig 8. SWB-Tension energies (in kcal/mol) for one-bond-sharing-[$n$]adamantanes. Upper plot in red: cyclo-one-BS-[11]adamantane (left ring of adamantane units) and cyclo-one-BS-[24]adamantane (right), and the plot of the SWB-tension energy per adamantane unit versus $n^{-2}$.

Lower plot in blue: one-BS-[5]adamantane and the plot of the SWB-tension energy per unit versus $n^{-1}$. For structures, the color code as in Fig. 1.

Acyclic and ring chains have molecular formulas $C_{8n+2}H_{10n+6}$ and $C_{8n}H_{10n}$, respectively; the partitioned formulas are $C_{2n-2}(CH)_{2n+2}(CH_2)4_{n+2}$ and $C_{2n}(CH)_{2n}(CH_2)_{4n}$, respectively. The total energy increases by 1116.7 eV for each new $C_8H_{10}$ or $C_2(CH)_2(CH_2)_4$ adamantane unit in both cases. MM2-calculated SWB-tension energy values increase by 24.4 kcal/mol for each new adamantane unit added to the acyclic chains with agreement between the intercepts (the $n \to \infty$ asymptotic SWB-tension energy per adamantane unit) for the two converging straight lines of Fig. 8. The minimal SWB-tension occurs around $n = 10$.

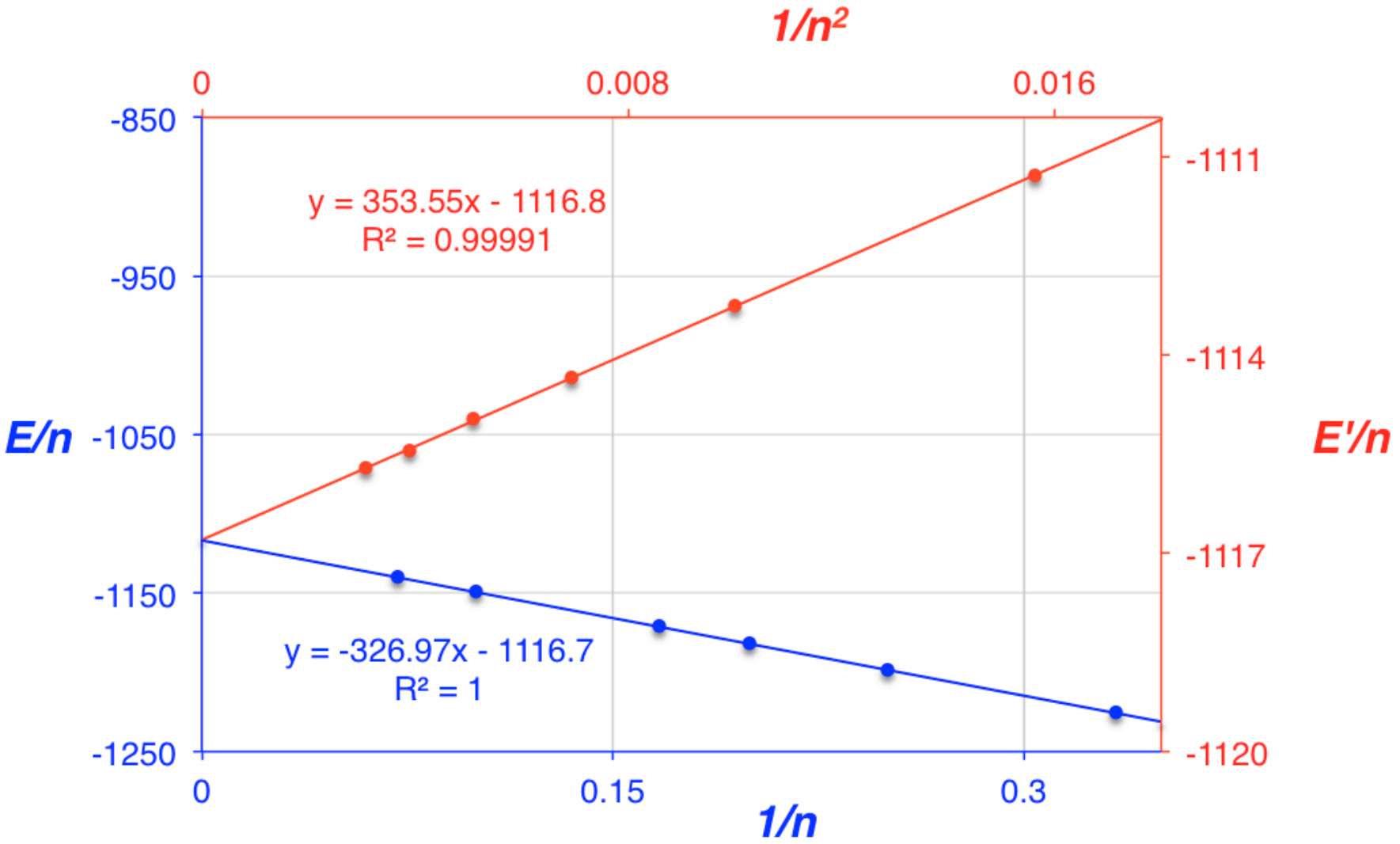

Fig. 9. Total energies (in eV) for one-bond-sharing-[$n$]adamantanes. Upper plot in red: cyclo-one-bond-sharing-[n]adamantanes and the plot of the total energy per diamantane unit $En^{-1}$ versus $n^{-2}$. Lower plot in blue: one-bond-sharing-[$n$]adamantanes and the plot of the total energy per unit versus $n^{-1}$.

### 3.4 One-bond-sharing-[*n*]diamantanes

For diamantane units, there are three pairs of $CH_2$–CH bonds at largest distances from each other, so that acyclic chains of one-bond-shared-[$n$]diamantanes are uniquely defined and form a linear structure as seen at the bottom of Fig. 7. The shared edge has two quaternary carbon atoms which, according to the color code, are the only red carbon atoms in the aggregate. The molecular formula for acyclic chains with $n$ diamantane units is $C_{12n+2}H_{14n+6}$ and the partitioned formula is $C_{2n-2}(CH)_{6n+2}$ $(CH_2)_{4n+6}$, whereas for cyclic chains these formulas are $C_{12n}H_{14n}$ , and $C_{2n}(CH)_{6n}(CH2)_{4n}$, respectively. In this class of diamondoid aggregates, the ring closure of such an acyclic aggregate without twisting may occur in two ways: the shared edges can be 'radial', i. e. orthogonal to the central symmetry axis of the large molecular circle, approximately like spokes of a wheel, or 'axial', i. e. parallel to the symmetry axis, as in Fig. 10. Molecular models show that the SWB-tension is lower when the ring closure occurs in the latter fashion: the diamantane units do not change their geometry appreciably, because the SWB-tension occurs with adjacent diamantane units pivoting around the shared bonds with both endpoints being quaternary carbon atoms.

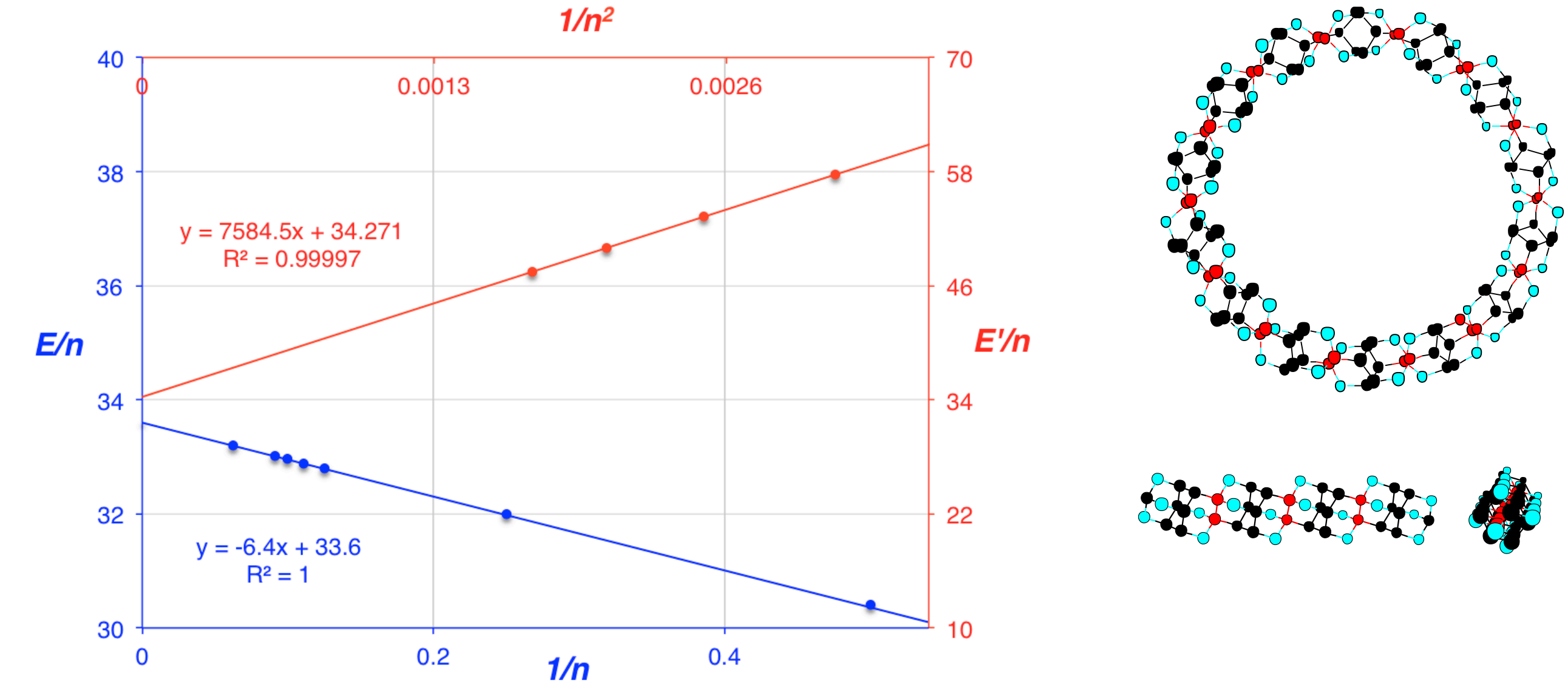

Fig 10. SWB-Tension energies (in kcal/mol) for one-bond-sharing-[$n$]diamantanes. Upper plot in red: cyclo-one-BS-[14]diamantane (axial), and the plot of the SWB-tension energy per diamantane unit versus $n^{-2}$. Lower plot in blue: one-BS-[4]diamantane (side and front views), and the plot of the SWB-tension energy per unit versus $n^{-1}$. For structures, the color code is as in Fig. 1.

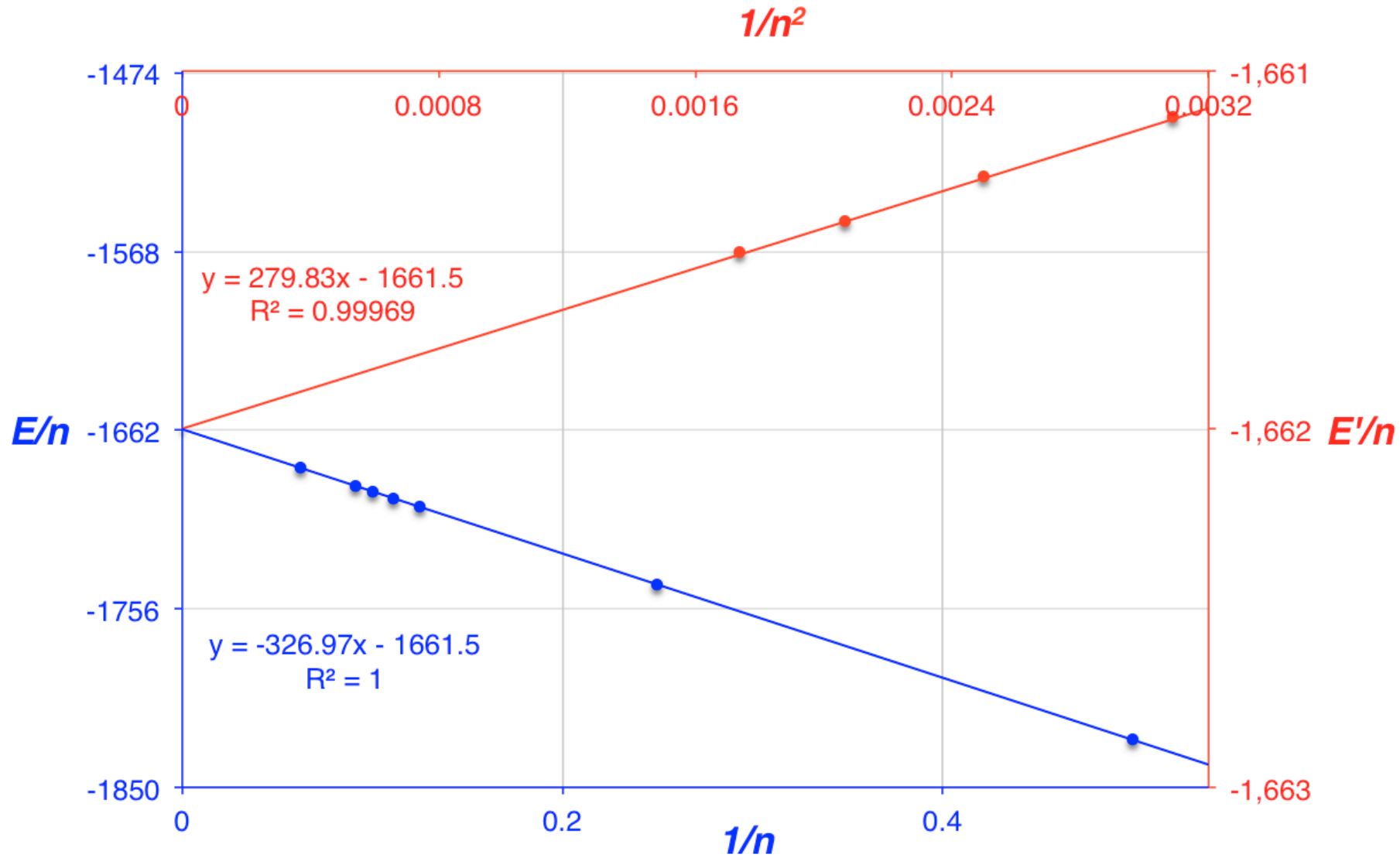

Fig. 11. Total energies (in eV) for one-bond-shared-[$n$]diamantanes with axial-bond quaternary carbon atoms. Upper plot in red: cyclo-one-bond-sharing-[$n$]diamantanes and the plot of the total

energy per diamantane unit $En^{-1}$ versus $n^{-2}$. Lower plot in blue: one-bond-sharing-[$n$]diamantanes and the plot of the total energy per unit versus $n^{-1}$.

### 3.5. Double-edge sharing systems

We decided not to include two-edge-shared systems, despite the fact that they had been mentioned in ref[22] because (i) they have never been approached synthetically, and (ii) for avoiding repulsion between hydrogen atoms on the "spine" of such acyclic systems (Fig. 13), the ground state conformation of acyclic chains is curved;  this latter fact adds complications to the cyclic structures sharing two "V-situated" edges with adjacent adamantane or diamantane units.

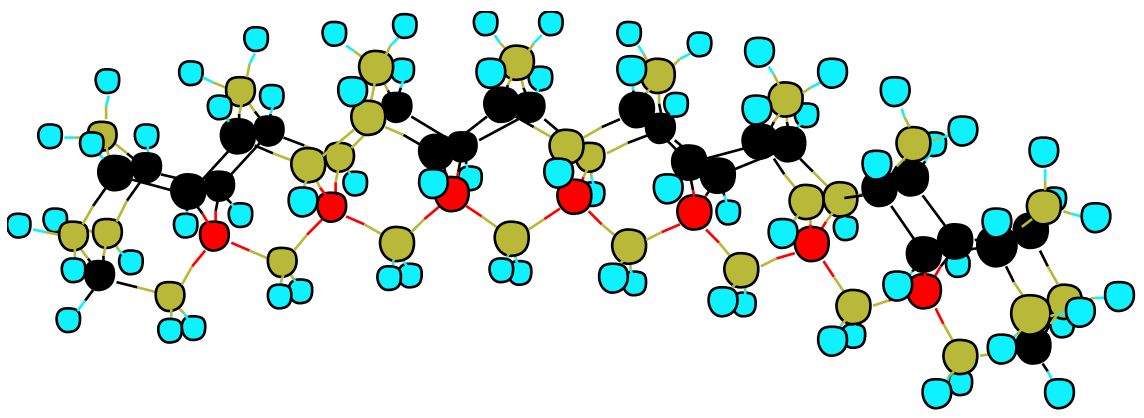

Fig. 13. Acyclic two-edge-sharing [8]adamantane showing hydrogen atoms (blue); by exception from other figures devoid of hydrogens, here hydrogens are shown in blue, whereas quaternary, tertiary and secondary carbon atoms are colored red, black, and olive, respectively.

### 3.6. Hexagon-sharing systems: [1234]-helix-*n*-diamantanes

Diamantane is the result of two adamantane units sharing a chair-shaped hexagon of carbon atoms, and higher catamantanes proceed along the same lines. In this last section we present a study of acyclic and cyclic helical cata-condensed aggregates of adamantane units sharing hexagons. It will be more convenient to consider these aggregates as resulting from associations between diamantane units conserving their reciprocal orientation, instead of adamantane units.

In Fig. 14 one can see a small portion of such an acyclic helix (in front and side views) with the dualist. Four adamantane units, or two diamantane units, are needed for one turn of the helix, as one can see in the middle front view of Fig. 14. The IUPAC (von Baeyer) nomenclature of such aggregates was discussed earlier.[21] The Balaban-Schleyer notation for catamantanes is based on dualists with digits from 1 to 4 in square brackets to indicate the four possible directions of bonds around points symbolizing $sp^3$-hybridized carbon atoms. One of the three isomeric tetramantanes, the chiral [123]tetramantane, is the precursor of the chiral helical systems discussed here. The cyclic [1234]-helix-*n*-diamantanes have molecular formulas $C_{16n}H_{16n}$ and partitioned molecular formulas $(CH_2)_{4n}(CH)_{8n}C_{4n}$, where $n$ is the number of hexagon-sharing diamantane units. In Figures 15 and 16 one can see the variation of the SWB-tension and total energy per diamantane unit versus $1/n^2$ for cyclic aggregates, and versus $1/n$ for acyclic aggregates, respectively.

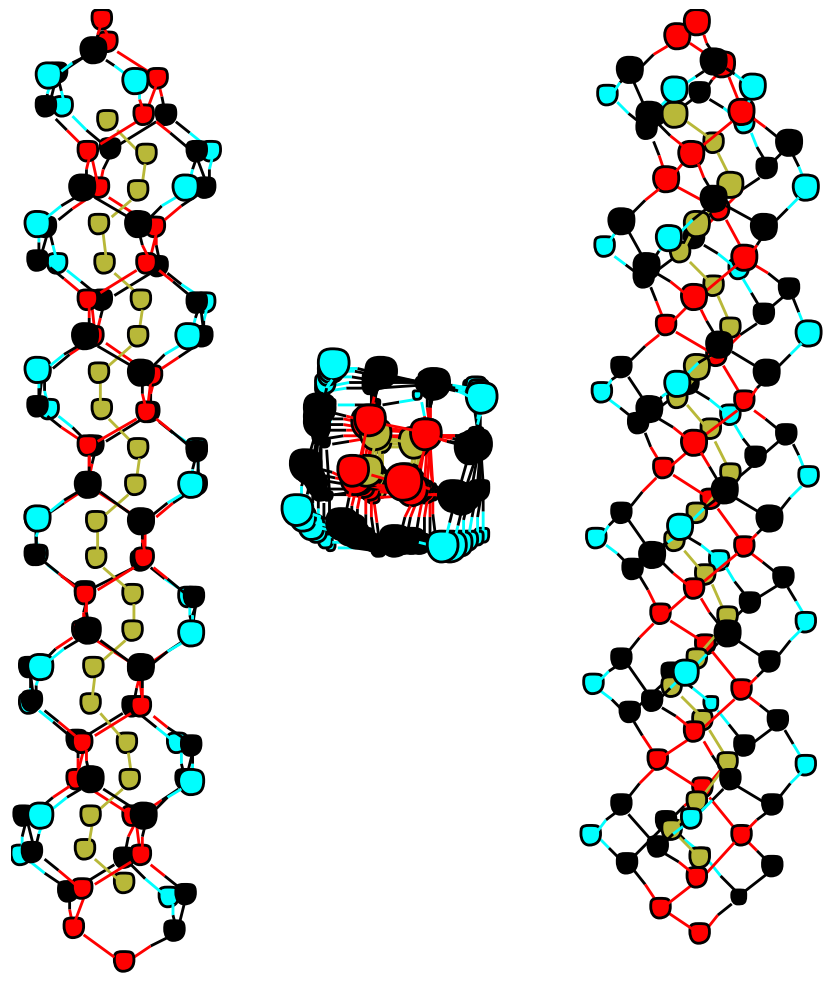

Fig. 14. Three views of a small fragment of acyclic [1234]-helix- *n*-diamantane. Its dualist is in olive, and carbon atoms have colors as in Fig. 1.

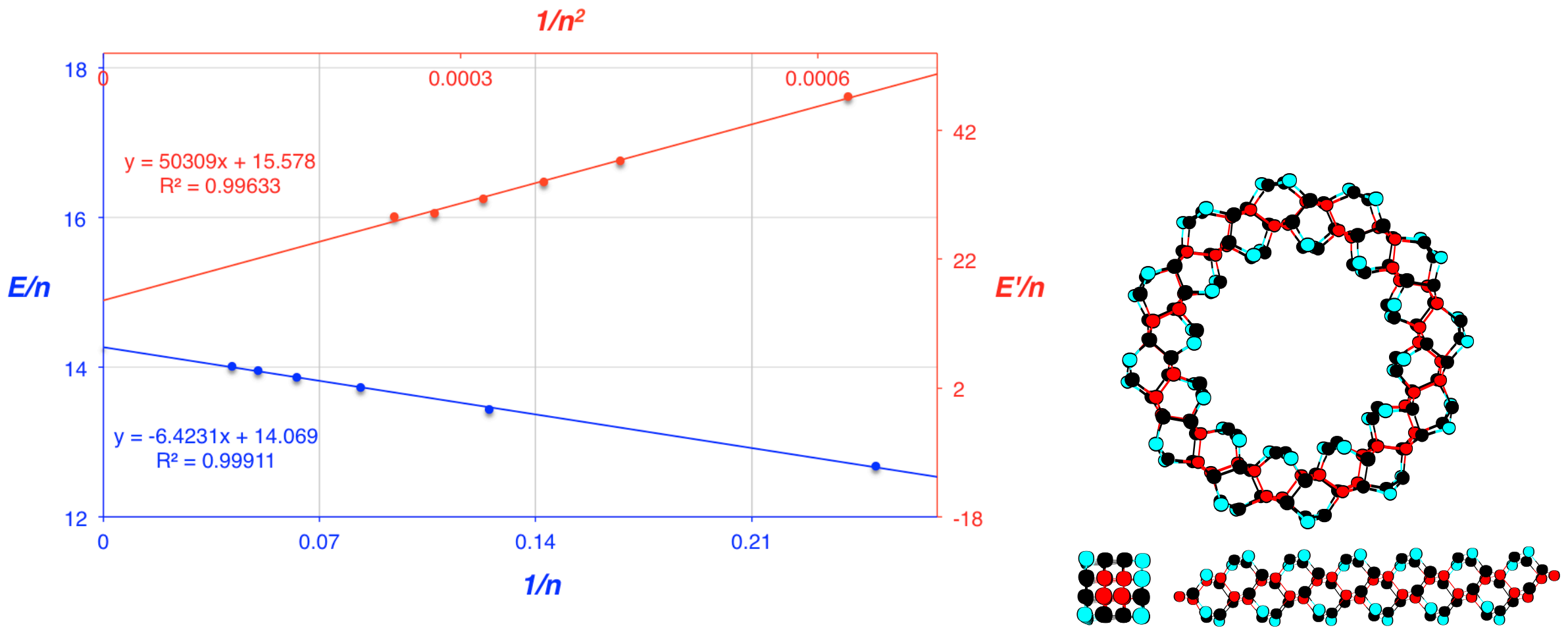

Fig 15. SWB-Tension energies (in kcal/mol) for [1234]-helix-*n*-adamantanes, exemplified by cyclo[1234]-helix-10-diamantane, and acyclic [1234]-helix-6-diamantane (front and side views), Upper plot in red: plot of the SWB-tension energy per adamantane unit versus $n^{-2}$ for cyclo[1234]-helix-*n*-adamantanes. Lower plot in blue: plot of the SWB-tension energy per adamantane unit versus $n^{-1}$ for acyclic [1234]-helix-*n*-adamantanes. For structures, the color code is as in Fig. 1.

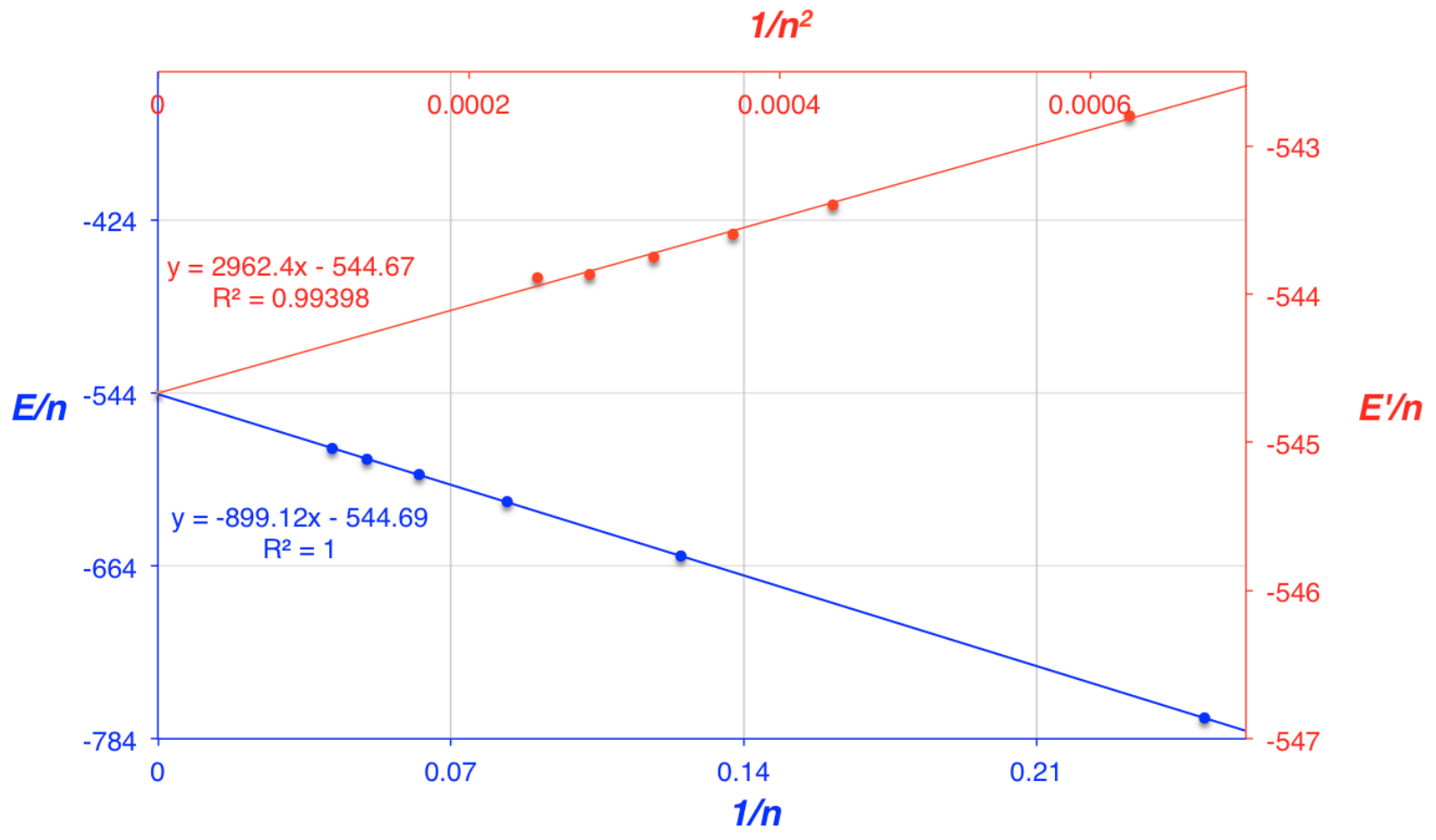

Fig 16. Total energies (in eV) for [1234]-helix-[*n*]adamantanes. Upper plot in red: total energy per adamantane unit versus $n^{-2}$ for cyclo[1234]-helix-[*n*]adamantanes. Lower plot in blue: total energy per unit versus $n^{-1}$ for acyclic [1234]-helix-[*n*]adamantanes.

In this class of adamantane chains, carbon atoms are part of the diamond crystalline lattice, unlike all other previous classes. For an acyclic chain having *n* adamantane units, the molecular formula is $C_{4n+6}H4_{n+12}$; the partitioned formula is $C_{n-2}(CH)_{2n+4}(CH_2)_{n+4}$. For cyclic chains, the molecular and partitioned formulas are $C_{4n}H_{4n}$ and $C_n(CH)_{2n}(CH_2)_n$, respectively. It would be possible to consider diamantane building blocks as the unit cells, but then each cell would consist of two adamantane units.

## 4. Conclusions

We have found reasonable approximate forms for the energies (both total and SWB-tension energies) of five classes of adamantane-based polymers. The forms for cyclic and acyclic chains appear slightly different. Plots of the SWB-tension energy per unit ($E/n$) versus $n^{-2}$ for cyclic aggregates reveal a good asymptotic linear correlation – the fits yield $R$-values only very slightly less than for the acyclic cases (see Tables 2 and 3). For three of these five classes of diamondoid aggregates, the straight-line fits of acyclic and cyclic plots meet satisfactorily at an asymptotic point $n \to \infty$. In all five classes, with increasing *n* values, the ratio *E/n* decreases for cyclic aggregates, but increases for acyclic aggregates.

Table 2. Linear correlations for total energies *E* (in eV) with correlation coefficient $R^2 \cong 1$.

| Species | Total Energy-Acyclic | Total Energy-Cyclic |
|---|---|---|
| spiro-[*n*]adamantanes (*c* = 9) | *E* = –1266.4 *n* – 177.4 | *E'* = –1266.4 *n* –167.5 |
| spiro-[*n*]diamantanes (*c* = 13) | *E* = –1811.1 *n* – 177.3 | *E* ' = –1811.2 *n* + 155.0 |
| one-BS-[*n*]adamantanes (*c* = 8) | *E* = –1116.7 *n* – 327.0 | *E*' = –1116.8 *n* + 353.6 |
| one-BS-[*n*]diamantanes (*c* = 12) | *E* = –1661.5 *n* – 327.0 | *E'* = –1661.5 *n* + 279.8 |
| [1234]-helix-[*n*]adamantanes (*c* = 4) | *E* = –544.7 *n* – 899.1 | *E'* = –544.9 *n* + 2692.4 |

SWB-Tension energies (in kcal/mol) computed with the MM2 program also afford good linear correlations with the number *n* of diamondoid units for acyclic aggregates, as indicated in Table 3. The equations in this table are another way of expressing the good linear correlations written in the lower part of Figures 4 through 16. The similarity of the monomer unit energies for the acyclic and cyclic systems cases is noted – the difference gives a possible upper bound on the error (for the given type of total energy computation). Table 3 gives results for the SWB tension energy, for the acyclic species, such as are judged to be more accurate asymptotic fits.

Table 3. Linear correlations for MM2-computed SWB-tension energies *E* (in kcal/mol) of acyclic diamondoid aggregates vs. *n*

| Acyclic species | SWB-tension energy |
|---|---|
| spiro-[*n*]adamantanes (*c*/cell = 9) | *E* = 27.7 *n* – 10.8 |
| spiro-[*n*]diamantanes (*c*/cell = 13) | *E* = 36.6 *n* – 9.5 |
| one-bond sharing [*n*]adamantanes (*c*/cell = 8) | *E* = 24.5 *n* – 7.3 |
| one-bond sharing [*n*]diamantanes (*c*/cell = 12) | *E* = 33.6 *n* – 6.4 |
| [1234]-helix-[*n*]adamantanes (*c*/cell = 4) | *E* = 14.1 *n* – 8.4 |

Equivalently one can formulate the energies in terms of the respective counts *c* and *h* of carbon and hydrogen atoms in a polymer, with the h/c ratio varying over the five classes of diamondoid chains. This equivalent fit for the equation

$$E_{PM6} = ac + bh \tag{6}$$

is given in Table 4 for the total energies of acyclic species. Yet another equivalent form would fit energies in terms of C–H and C–C bonds.

With the energy conversion factor 1 eV = 23.06 kcal/mol, it is easy to see that the "SWB-tension energy" is smaller than the total energy by two orders of magnitude (therefore the need for 4 or 5 significant digits for the latter, if one wishes to detect steric tension).

Table 4. Correlations between the total energy for acyclic diamondoid chains and the numbers *c* and *h* of their carbon and hydrogen atoms shown in Tables S1, S3, S5, S7, S9.

| Acyclic structure | *c* | *h* | *a* | *b* |
|---|---|---|---|---|
| spiro-[*n*]adamantanes | 9*n*+1 | 14*n*+2 | –122.21 | –13.87 |
| spiro-[*n*]diamantanes | 13*n*+1 | 18*n*+2 | –122.27 | –13.85 |
| one-bond sharing [*n*]adamantanes | 8*n*+2 | 10*n*+6 | –122.51 | –13.66 |
| one-bond sharing [*n*]diamantanes | 12*n*+2 | 14*n*+6 | –122.53 | –13.65 |
| [1234]helix [*n*]adamantanes | 4*n*+6 | 4*n*+12 | –122.49 | –13.69 |

Supplemental information about values of PM6-computed total energies and MM2-computed SWB-tension energies (with details about the constituent parts of these SWB-tension energies) is provided for selected *n* values in Tables S1 to S10. For cyclic aggregates the row colored in red indicates the minimal SWB-tension energy. Table 5 provides information on the number of units in the cyclic chain with minimal SWB-tension energy for each class of diamondoids investigated in the present paper.

Table 5. Cyclic diamondoid chains with minimal SWB-tension energy

| Cyclic species | $n_{min}$ via eq. (5) | $n_{min}$ (observed) |
|---|---|---|
| spiro-[*n*]adamantanes | 10 | 10 |
| spiro-[*n*]diamantanes | 10 | ≤ 12 |
| one-bond sharing-[*n*]adamantanes | 17 | ≥ 14 |
| one-bond sharing-[*n*]diamantanes | 15 | ≤ 18 |
| [1234]helix-[*n*]adamantanes | 59 | ≤ 60 |

In a future paper on cyclic and acyclic aggregates of hexagon-sharing diamantane units, it is to be investigated whether for regular and irregular catamantanes as well as for perimantanes that some of these aggregates behave in more complicated ways than the aggregates examined in this communication. The present data may offer yet another way of comparing the strain stiffness for diamond nanorods and single-walled nanotubes. Published data indicate comparable strength and strain stiffness.[41-44]

**Acknowledgement.** The authors acknowledge the support of the Welch Foundation, through grant BD-0894.

**Supplements**

The following Tables provide results of calculations using the MM2 program for strain energies (in kcal/mol), and the PM6 program for total energies (in eV); 1 eV = 23 kcal/mol. The row in red for cyclics corresponds to the lowest strain energy in the corresponding Table.

Table S1. Strain energies (MM2, in kcal/mol) for acyclic spiro-[*n*]adamantanes

| *n* | Total Energy | Stretch | Bend | Stretch-Bend | Torsion | Non-1,4 VDW | 1,4 VDW | SWB-Strain Energy |
|---|---|---|---|---|---|---|---|---|
| 4 | -5242.8 | 8.7 | 11.8 | 1.1 | 43.4 | -3.4 | 38.3 | 99.9 |
| 5 | -6509.3 | 11.2 | 15.7 | 1.5 | 55.1 | -3.7 | 47.8 | 127.6 |
| 6 | -7775.5 | 13.8 | 19.6 | 1.9 | 66.7 | -3.9 | 57.2 | 155.3 |
| 7 | -9041.7 | 16.3 | 23.5 | 2.2 | 78.4 | -4.2 | 66.7 | 182.9 |
| 8 | -10308.6 | 18.9 | 27.3 | 2.6 | 90.0 | -4.5 | 76.2 | 210.6 |
| 9 | -11574.2 | 21.5 | 31.2 | 2.9 | 101.7 | -4.8 | 85.7 | 238.3 |
| 10 | -12841.3 | 24.0 | 35.1 | 3.3 | 113.4 | -5.0 | 95.2 | 265.9 |

Table S2. Spiro-cyclo[*n*]adamantanes

| *n* | PM6 Energy | Stretch | Bend | Stretch-Bend | Torsion | Non-1,4 VDW | 1,4 VDW | SWB-Strain Energy |
|---|---|---|---|---|---|---|---|---|
| 6 | -7570.5 | 83.5 | 266.1 | 13.9 | 100.0 | 108.7 | 114.2 | 686.3 |
| 7 | -8841.2 | 64.0 | 233.7 | 10.6 | 111.6 | 92.7 | 117.3 | 629.7 |
| 8 | -10110.7 | 55.5 | 212.0 | 8.9 | 122.0 | 81.8 | 121.0 | 601.2 |
| 9 | -11379.5 | 51.8 | 196.9 | 8.0 | 132.0 | 74.1 | 125.6 | 588.4 |
| 10 | -12647.8 | 50.2 | 185.8 | 7.5 | 141.9 | 68.4 | 131.1 | 584.9 |
| 11 | -13915.8 | 49.9 | 177.6 | 7.3 | 151.9 | 63.6 | 137.2 | 587.5 |
| 12 | -15183.4 | 50.3 | 171.3 | 7.3 | 162.0 | 59.5 | 143.9 | 594.3 |
| 13 | -16450.9 | 51.1 | 166.5 | 7.3 | 172.2 | 56.0 | 151.1 | 604.2 |
| 14 | -17718.2 | 52.3 | 162.9 | 7.4 | 182.6 | 52.6 | 158.7 | 616.5 |
| 16 | -20252.4 | 55.2 | 158.4 | 7.7 | 203.8 | 46.8 | 174.5 | 646.3 |
| 18 | -22786.3 | 58.7 | 156.4 | 8.1 | 225.4 | 41.7 | 191.1 | 681.4 |

Table S3. Acyclic spiro-[*n*]diamantanes

| *n* | Total Energy | Stretch | Bend | Stretch-Bend | Torsion | Non-1,4 VDW | 1,4 VDW | SWB-Strain Energy |
|---|---|---|---|---|---|---|---|---|
| 2 | -3799.6 | 4.6 | 4.2 | 0.5 | 32.8 | -7.6 | 29.2 | 63.6 |
| 3 | -5610.7 | 7.7 | 7.9 | 0.8 | 50.7 | -10.5 | 43.7 | 100.2 |
| 4 | -7421.8 | 10.7 | 11.5 | 1.2 | 68.6 | -13.5 | 58.1 | 136.7 |
| 5 | -9233.0 | 13.8 | 15.2 | 1.6 | 86.6 | -16.5 | 72.6 | 173.3 |
| 6 | -11044.1 | 16.8 | 18.9 | 2.0 | 104.5 | -19.5 | 87.1 | 209.9 |
| 7 | -12855.2 | 19.9 | 22.6 | 2.4 | 122.5 | -22.5 | 101.6 | 246.4 |
| 8 | -14666.4 | 22.9 | 26.3 | 2.7 | 140.4 | -25.5 | 116.1 | 283.0 |

Table S4. Cyclo-spiro[*n*]diamantanes

| Species | Total Energy | Stretch | Bend | Stretch-Bend | Torsion | Non-1,4 VDW | 1,4 VDW | SWB-Strain Energy |
|---|---|---|---|---|---|---|---|---|
| 12 | -21721.0 | 61.4 | 169.6 | 9.5 | 233.5 | 24.9 | 205.9 | 704.7 |
| 14 | -25345.4 | 63.2 | 158.9 | 9.3 | 267.0 | 11.7 | 230.4 | 740.6 |
| 16 | -28968.6 | 66.3 | 152.6 | 9.5 | 301.1 | 0.1 | 256.0 | 785.6 |
| 18 | -32592.5 | 70.1 | 149.2 | 9.7 | 335.6 | -10.4 | 282.3 | 836.5 |
| 20 | -36215.7 | 74.4 | 147.9 | 10.1 | 370.4 | -20.3 | 309.2 | 891.7 |
| 22 | -39838.6 | 79.0 | 148.1 | 10.6 | 405.3 | -29.6 | 336.4 | 949.9 |
| 24 | -43461.4 | 84.0 | 149.7 | 11.1 | 440.4 | -38.7 | 364.0 | 1010.6 |

Table S5. Acyclic one-bond-sharing-[*n*]adamantanes

| *n* | Total Energy | Stretch | Bend | Stretch-Bend | Torsion | Non-1,4 VDW | 1,4 VDW | SWB-Strain Energy |
|---|---|---|---|---|---|---|---|---|
| 3 | -3677.0 | 5.1 | 6.8 | 0.6 | 30.1 | -3.1 | 26.6 | 66.1 |
| 4 | -4793.7 | 7.2 | 10.3 | 0.8 | 40.9 | -3.4 | 34.8 | 90.6 |
| 5 | -5910.4 | 9.2 | 13.8 | 1.1 | 51.7 | -3.7 | 43.1 | 115.1 |
| 6 | -7027.0 | 11.2 | 17.2 | 1.3 | 62.4 | -4.0 | 51.3 | 139.5 |
| 10 | -11493.7 | 19.4 | 31.1 | 2.3 | 105.5 | -5.2 | 84.3 | 237.5 |
| 14 | -15960.5 | 27.5 | 45.0 | 3.4 | 148.6 | -6.4 | 117.3 | 335.4 |

Table S6. Cyclic one-bond-sharing-[*n*]adamantanes

| *n* | Total Energy | Stretch | Bend | Stretch-Bend | Torsion | Non-1,4 VDW | 1,4 VDW | SWB-Strain Energy |
|---|---|---|---|---|---|---|---|---|
| 8 | -8890.2 | 165.9 | 445.2 | 29.3 | 131.8 | 140.5 | 165.5 | 1078.3 |
| 10 | -11132.6 | 125.9 | 377.5 | 22.5 | 149.9 | 130.5 | 168.2 | 974.6 |
| 12 | -13372.0 | 104.1 | 287.1 | 19.1 | 166.7 | 120.7 | 164.8 | 862.6 |
| 14 | -15609.6 | 92.1 | 251.6 | 16.1 | 184.6 | 110.2 | 172.3 | 826.8 |
| 16 | -17847.0 | 79.5 | 275.8 | 11.5 | 201.5 | 95.9 | 187.1 | 851.3 |
| 18 | -20082.5 | 78.5 | 266.2 | 11.1 | 221.1 | 86.7 | 199.9 | 863.4 |

Table S7. Acyclic one-bond-sharing-[*n*]diamantanes

| *n* | Total Energy | Stretch | Bend | Stretch-Bend | Torsion | Non-1,4 VDW | 1,4 VDW | SWB-Strain Energy |
|---|---|---|---|---|---|---|---|---|
| 2 | -3649.9 | 4.0 | 3.6 | 0.4 | 31.9 | -7.3 | 28.2 | 60.8 |
| 4 | -6972.8 | 8.8 | 9.9 | 0.9 | 66.3 | -13.3 | 55.4 | 128.0 |
| 8 | -13618.6 | 18.2 | 22.5 | 2.0 | 134.9 | -25.1 | 109.8 | 262.4 |
| 9 | -15280.1 | 20.6 | 25.6 | 2.2 | 152.1 | -28.0 | 123.5 | 296.0 |
| 10 | -16941.5 | 23.0 | 28.8 | 2.5 | 169.2 | -31.0 | 137.1 | 329.6 |
| 11 | -18603.0 | 25.4 | 31.9 | 2.8 | 186.4 | -33.9 | 150.7 | 363.2 |
| 16 | -26910.3 | 37.2 | 47.6 | 4.1 | 272.2 | -48.7 | 218.7 | 531.2 |

Table S8. Cyclic one-bond-shared-[*n*]diamantanes with axial quaternary carbon atoms.

| n | Total Energy | Stretch | Bend | Stretch-Bend | Torsion | Non-1,4 VDW | 1,4 VDW | Strain Energy |
|---|---|---|---|---|---|---|---|---|
| 18 | -29891.3 | 88.0 | 265.1 | 15.0 | 323.9 | 42.2 | 303.6 | 1037.9 |
| 20 | -33215.9 | 89.9 | 245.8 | 15.4 | 353.4 | 35.6 | 325.1 | 1065.2 |
| 22 | -36540.2 | 90.6 | 236.2 | 15.1 | 386.6 | 23.3 | 347.2 | 1099.0 |
| 24 | -39864.1 | 92.1 | 229.2 | 14.9 | 420.0 | 11.6 | 370.2 | 1138.1 |
| 26 | -43102.2 | 94.5 | 224.2 | 14.9 | 453.6 | 0.3 | 393.8 | 1181.3 |

Table S9. Acyclic [1234]-helix-*n*-adamantane

| *n* | Total Energy | Stretch | Bend | Stretch -Bend | Torsion | Non-1,4 VDW | 1,4 VDW | SWB-Strain Energy |
|---|---|---|---|---|---|---|---|---|
| 4 | -3077.9 | 3.0 | 1.9 | 0.2 | 28.7 | -8.2 | 24.4 | 49.9 |
| 8 | -5256.7 | 7.9 | 7.1 | 0.6 | 58.8 | -12.3 | 43.9 | 105.9 |
| 12 | -7435.5 | 12.5 | 12.7 | 0.9 | 88.9 | -16.1 | 63.4 | 162.4 |
| 16 | -9614.2 | 17.1 | 18.4 | 1.3 | 119.0 | -20.0 | 82.9 | 218.7 |
| 20 | -11793.0 | 21.7 | 24.0 | 1.6 | 149.1 | -23.9 | 102.5 | 275.1 |
| 24 | -13971.8 | 26.4 | 29.6 | 2.0 | 179.3 | -27.7 | 122.0 | 331.5 |

Table S10. Cyclic [1234]-helix-*n*-adamantane

| *n* | Total energy | Stretch | Bend | Stretch-Bend | Torsion | Non-1,4 VDW | 1,4 VDW | SWB-Strain Energy |
|---|---|---|---|---|---|---|---|---|
| 40 | -21712 | 218.6 | 740 | 31.8 | 353.4 | 196.9 | 348.3 | 1889,0 |
| 48 | -26083 | 178.8 | 651.7 | 24.3 | 406.8 | 164.5 | 365.1 | 1791.2 |
| 52 | -28267 | 168.2 | 617.5 | 21.9 | 433.9 | 150.2 | 375.4 | 1767.1 |
| 56 | -30450 | 161 | 588.3 | 20.1 | 461.4 | 136.9 | 386.7 | 1754.3 |
| 60 | -32632 | 156.3 | 563.4 | 18.7 | 489.1 | 124.2 | 399 | 1750.7 |
| 64 | -34809 | 170.1 | 573.2 | 18 | 519 | 117.8 | 433.5 | 1831.6 |

Graphical Abstract

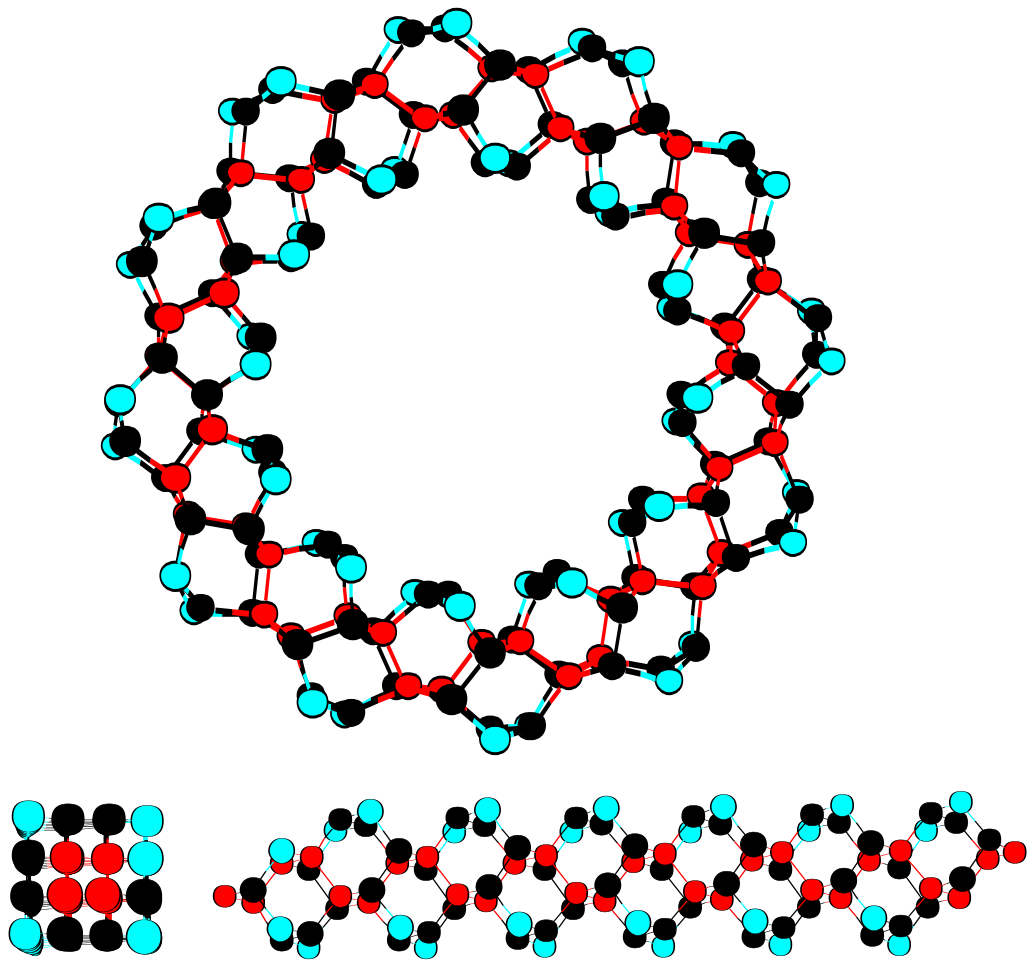